\documentclass{article}
\usepackage{amssymb}
\usepackage{amsmath}
\usepackage{graphicx}
\usepackage{epsfig}
\usepackage{amsfonts}
\usepackage{bm}
\begin{document}
\begin{center}
Quantum Simulation of Markov Chains \vskip 0.5cm X. F. Liu\footnote[1]{email
address: liuxf@pku.edu.cn} \\ Department of Mathematics and LMAM\\ Peking University, Beijing 100871, China\\

\end{center}
\vskip 0.5cm

\begin{flushleft}
Abstract. The possibility of simulating a stochastic process by the intrinsic randomness of
quantum system is investigated. Two simulations of Markov Chains by the measurements of
quantum systems are proposed.
\end{flushleft}
\section{Introduction}
Stochastic simulation methods, also commonly known as Monte Carlo methods, are
important in scientific computing. Such methods are useful for studying
deterministic problems that are too complicated to model analytically or
deterministic problems whose high dimensionality makes standard discretizations
infeasible. Stochastic simulations [1] are usually realized by classical computer
algorithms, which are actually deterministic. In this paper, we study the
possibility of simulating a stochastic process by the intrinsic randomness of
quantum system. We believe that, theoretically speaking,  a quantum system may simulate a stochastic process better
than a classical system due to its intrinsic randomness. 

Markov Chain is a basic and widely studied stochastic process. It is also
useful in scientific computing. For example, one can solve partial differential
equations or groups of linear equations by simulations of Markov Chains. In
this paper we will propose two simulations of Markov Chains by the measurements of
quantum systems. The second simulation is closely related to the reading of a
quantum register and thus may be realized by quantum computer. It thus seems that quantum computer might be a better choice than classical computer for
carrying out Monte Carlo method.
\section{Definition of Markov Chain}
In probability theory, the concept of experiment occupies a crucial position.
Roughly speaking, an experiment consists of a space of its possible outcomes,
together with an assignment of probabilities to each of these outcomes. In this
paper, we only consider experiments with finitely many outcomes. Let $I$ be a
finite set. The formal definition of an experiment (with finitely many
outcomes)is as follows[2].

Definition 1. A function $\lambda:I\rightarrow [0,1]$ is called a distribution
on $I$ if $\sum_{i\in I}\lambda(i)=1$. When $\lambda$ is a distribution on $I$
it will be denoted by $\lambda=(\lambda_i:i\in I)$ where
$\lambda_i=\lambda(i)$.

Definition 2. An experiment $E$ is a pair $(\Omega,p)$ where
$\Omega=\{\omega_i|i\in I\}$ is a finite set of the outcomes of $E$, called the
sample space of $E$, and $p=(p_i:i\in I)$ is a distribution on $I$, called the
distribution of $E$, where $p_i=p(\omega_i)$.

Another ingredient necessary for the definition of Markov Chain is the concept
of stochastic matrix.

Definition 3. A matrix $P=(p_{ij})_{i,j\in I}$ is called stochastic if for each
$i\in I$ $p^i=(p_{ij}:j\in I)$ is a distribution on $I$.

Now let us introduce the definition of Markov Chain.

Definition 4. Let $P=(p_{ij})_{i,j\in I}$ be a stochastic matrix. A (finite or
infinite sequence) of experiments $(E_n)_{n\geq 0}$ is called a Markov Chain
with the transition matrix $P$ if all the experiments in the sequence have the
same sample space $\Omega=\{\omega_i|i\in I\}$ and for each $n\geq 0$ the
distribution of the experiment $E_{n+1}$ only depends on the outcome of the
experiment $E_n$ in the following way: when the outcome of $E_n$ is $\omega_i$
the distribution of $E_{n+1}$ is $p^i=(p_{ij}:j\in I)$.

From probability point of view, the behavior of a Markov Chain $(E_n)_{n\geq
0}$ is completely determined by its transition matrix and its initial
distribution, namely, the distribution of the experiment $E_0$.

Remark. The above definition of Markov Chain might not be formal or rigorous
enough from mathematical point of view. But for our modest purpose in this
paper Definition 4 is adequate. For the definition of Markov Chain based on the
terminology of probability space and conditional probability [3], which is
mathematically beyond question, the reader who has a passion for mathematical
rigorousness and preciseness can refer to mathematical text books on
probability theory.
\section{Simulation of Markov Chains by Angular Momentum System}
In this section we will propose a realization of Markov Chains by a sequence of
quantum measurements for angular momentums.

Consider a spin $s$ system ${\mathbb S}$. Let $\hat{S}_x$, $\hat{S}_y$ and
$\hat{S}_z$ be the $x$, $y$ and $z$ components of the spin operator
${\bf\hat{S}}$ of ${\mathbb S}$.

Let $R$ be the operator of a rotation through an angle $\theta$ about a
direction specified by the unit vector ${\bf n}$. Then we have $R=e^{-i\theta
{\bf n}\cdot {\bf\hat{S}}}$, which is clearly unitary. Suppose that this
rotation is specified by the Eulerian angles $(\alpha,\beta,\gamma)$, then by
definition it can be carried out in three stages: (1) a rotation through the
angle $\alpha$ ($0\leq\alpha\leq 2\pi$) about the $z$-axis, (2) a rotation
through the angle $\beta$ ($0\leq\beta\leq\pi$) about the new position $y'$ of
the $y$-axis, (3) a rotation through the angle $\gamma$ ($0\leq\gamma\leq 2\pi$)
about the resulting final position $z''$ of the $z$-axis. Correspondingly, we
have
$$R=e^{-i\theta
{\bf n}\cdot
{\bf\hat{S}}}=e^{-i\gamma\hat{S}_{z''}}e^{-i\beta\hat{S}_{y'}}e^{-i\alpha\hat{S}_{z}}
=e^{-i\alpha\hat{S}_{z}}e^{-i\beta\hat{S}_{y}}e^{-i\gamma\hat{S}_{z}}.$$

Now define $\hat{S}_{\bf n}=R^{\dag}\hat{S}_zR$. Physically, it represents the
component of the spin operator ${\bf\hat{S}}$ in the direction specified by the
vector ${\bf n}$. Let $\{|sm\rangle|m=s,s-1,\cdots,-s\}$ be the state vectors
of the spin system ${\mathbb S}$ that span the standard $2s+1$ dimensional
irreducible representation space of the Lie algebra generated by the spin
operators $\hat{S}_x$, $\hat{S}_y$ and $\hat{S}_z$. To be precise, we have
$${\bf\hat{S}}^2|sm\rangle=s(s+1)|sm\rangle,\ \hat{S}_z|sm\rangle=m|sm\rangle.$$
Let $|sm'\rangle=R^{-1}|sm\rangle$ for each $m$. Then we have
$$\langle sm_2'|sm_1\rangle=\langle sm_2|e^{-i\alpha\hat{S}_{z}}e^{-i\beta\hat{S}_{y}}
e^{-i\gamma\hat{S}_{z}}|sm_1\rangle=e^{-im_2\alpha}\langle
sm_2|e^{-i\beta\hat{S}_{y}}|sm_1\rangle e^{-im_1\gamma}.$$ Denoting this
quantity by $D^s_{m_2m_1}(\alpha,\beta,\gamma)$, we have
$$D^s_{m_2m_1}(\alpha,\beta,\gamma)=e^{-im_2\alpha}d^s_{m_2m_1}(\beta)e^{-im_1\gamma},$$
where
$$d^s_{m_2m_1}(\beta)=\langle
sm_2|e^{-i\beta\hat{S}_{y}}|sm_1\rangle.$$

Similarly, it is easy to check that
$$\langle sm_2|sm_1'\rangle=\overline{D^s_{m_1m_2}(\alpha,\beta,\gamma)}
=e^{im_1\alpha}\overline{d^s_{m_1m_2}(\beta)}e^{im_2\gamma},$$ where the
``overline" means taking complex conjugation. But it can be proved that
$$\overline{d^s_{m_1m_2}(\beta)}=d^s_{m_1m_2}(\beta)=(-)^{m_1-m_2}d^s_{m_2m_1}(\beta),$$
it thus follows that
$$\left |\langle sm_2'|sm_1\rangle\right |^2=\left |\langle sm_2|sm_1'\rangle\right |^2
=\left |d^s_{m_2m_1}(\beta)\right |^2.$$

Notice that when the system ${\mathbb S}$ is in the state $|sm'\rangle$ and a
measurement of $S_z$ is carried out then $\left |\langle
sm_2|sm_1'\rangle\right |^2$ is exactly the probability that the measurement
will give the value $m_2$. Likewise, when the system ${\mathbb S}$ is in the
state $|sm\rangle$ and a measurement of $S_{\bf n}$ is carried out then $\left
|\langle sm_2'|sm_1\rangle\right |^2$ is exactly the probability that the
measurement will give the value $m_2$. That these two probabilities are
identical, as shown above, prompts us to propose the following realization of
Markov Chains by the proper measurements on the system ${\mathbb S}$.

Denote by $M$ and $M'$ respectively the measurements of $S_z$ and $S_{\bf n}$
on the system ${\mathbb S}$. Let us consider the sequence $(M_n)_{n\geq 0}$ of
measurements, where $M_n$ stands for $M$ when $n$ is even and stands for $M'$
when $n$ is odd. We observe that each $M_n$ can naturally be regarded as an
experiment in the sense of Definition 2, with the sample space
$\Omega=\{s,s-1,\cdots,-s\}\triangleq I$, which is independent of $n$. On the
other hand, the distribution of $M_{n+1}$ only depends on the outcome of
$M_{n}$ in the following way: when the outcome of $M_{n}$ is $m_1$, the
distribution of $M_{n+1}$ is $p^{m_1}=(|d^s_{m_2m_1}(\beta)|^2:m_2\in I)$. The
reason is, if the measurement $M$ or $M'$ gives the value $m_1$, then the
system will collapse to the state $|sm_1\rangle$ or $|sm_1'\rangle$ after the
measurement.

Now it should be clear that the above defined the sequence $(M_n)_{n\geq 0}$ of
measurements can be regarded as a simulation of the Markov Chain with the
transition matrix $(p_{ij})_{i,j\in I}$ where $p_{ij}=|d^s_{ji}(\beta)|^2$. If
the system ${\mathbb S}$ is initially in the state $|\psi\rangle$, then this
Markov Chain has the initial distribution $\{\langle sm|\psi\rangle|^2:m\in
I\}$.

An interesting case arise when we take $s=1/2$, $\beta=\pi/2$ and
$\alpha=\gamma=0$. In this case ${\bf\hat{S}_n}=\hat{S}_x$ and the sample space
of $M_n$ is $\{1/2,-1/2\}$. Moreover, we have
$$|d^{1/2}_{m_2m_1}(\pi/2)|^2=1/2,\ \forall\, m_1,m_2\in\{1/2,-1/2\}.$$
Thus the sequence $(M_n)_{n\geq 0}$ can be regarded as a simulation of the coin
tossing experiment and the system ${\mathbb S}$ may be used as a random numbers
generator.

\section{Simulation of Markov Chains by q-Bit System}
Essentially based on the idea of the last section we propose in this section a
simulation of Markov Chains by q-bit system.

We model a q-bit as a spin $1/2$ system ${\mathbb S}$ and keep the same
notation for the system ${\mathbb S}$ as in the last section. Let us consider
the q-bit system ${\mathbb Q}_N$ that is composed of $N$ independent q-bits.
Denote by ${\bf\hat{S}}^i$ the angular momentum operator of the $i$th q-bit and
by ${\bf\hat{J}}$ the angular momentum operator of the q-bit system ${\mathbb
Q}_N$. As in the last section we define $\hat{J}_{\bf n}=R^{\dag}\hat{J}_zR$.
By definition we have
$${\bf\hat{J}}=\sum_{i=1}^N{\bf\hat{S}}^i,\  \hat{J}_{\bf n}=\sum_{i=1}^N
\hat{S}_{\bf n}^i.$$

Now consider the measurement of $J_z$ carried out by measuring each $S_z^i$ and
the measurement of $J_{\bf n}$ carried out by measuring each $S_{\bf n}^i$. We
denote these two kinds of measurement by $M$ and $M'$ respectively.

For convenience, we introduce the state vectors $|+\rangle$, $|-\rangle$,
$|+_{\bf n}\rangle$ and $|-_{\bf n}\rangle$ as follows:
$$S_z|+\rangle=\frac{1}{2}|+\rangle,S_z|-\rangle=-\frac{1}{2}|-\rangle,
S_{\bf n}|+_{\bf n}\rangle=\frac{1}{2}|+_{\bf n}\rangle, S_{\bf n}|-_{\bf
n}\rangle=-\frac{1}{2}|-_{\bf n}\rangle.$$Then, from the general formula for
the matrix element $d^s_{m_2m_1}(\beta)$ we have
$$|\langle+_{\bf n}|+\rangle|^2=|\langle+|+_{\bf n}\rangle|^2|=|\cos
\beta/2|^2,\ |\langle-_{\bf n}|+\rangle|^2=|\langle-|+_{\bf n}\rangle|^2=|\sin
\beta/2|^2$$ and $$|\langle+_{\bf n}|-\rangle|^2=|\langle+|-_{\bf n}\rangle|^2|
=|\sin \beta/2|^2,\ |\langle-_{\bf n}|-\rangle|^2=|\langle-|-_{\bf
n}\rangle|^2=|\cos \beta/2|^2.$$

According to the theory of quantum measurement, after the measurement $M$ each
q-bit is either in the state $|+\rangle$ or in the state $|-\rangle$ and after
the measurement $M'$ each q-bit is either in the state $|+_{\bf n}\rangle$ or
in the state $|-_{\bf n}\rangle$. Thus, after the measurement $M$, if its
outcome is $j$, then
 $N/2+j$ q-bits of the system ${\mathbb Q}_N$
collapse to the state $|+\rangle$ and the other $N/2-j$ q-bits collapse to the
state $|-\rangle$.

Denote by $q_{j'j}$ the probability that the measurement $M'$ on the system
${\mathbb Q}_N$ which has just experienced the measurement $M$ that gives the
value $j$, will give the value $j'$. It then follows that if $j\geq j'$
$$q_{j'j}=\sum_{m=j-j'}^{N/2-j'}{N/2+j \choose m}{N/2-j \choose N/2-j'-m}
|\cos \beta/2|^{2(N+j-j'-2m)}|\sin \beta/2|^{2(j'-j+2m)}$$  and if $j\leq j'$
$$q_{j'j}=\sum_{m=j'-j}^{N/2+j'}{N/2-j \choose m}{N/2+j \choose N/2+j'-m}
|\cos \beta/2|^{2(N-j+j'-2m)}|\sin \beta/2|^{2(j-j'+2m)}.$$ Moreover, it is
easily check that $q_{j'j}$ is also the probability that the measurement $M$ on
the system ${\mathbb Q}_N$ which has just experienced the measurement $M'$ that
gives the value $j$, will give the value $j'$. It should then be clear that the
sequence $(M_n)_{n\geq 0}$ of measurements, where $M_n$ stands for $M$ when $n$
is even and stands for $M'$ when $n$ is odd, can be regarded as a simulation of
the Markov Chain with the transition matrix $(q_{j'j})_{-N/2\leq j',j\leq
N/2}$.
\vskip 1cm
References
\begin{enumerate}
\item Brian D. Ripley, Stochastic Simulation, Wiley 1987.
\item J.R. Norris, Markov Chains, Cambridge University Press 1997.
\item Daniel W. Strook, An Introduction to Markov Processes, Springer 2005.
\end{enumerate}
\end{document}